\documentclass[conference]{IEEEtran}
\usepackage{amsmath,amssymb,amsfonts}
\usepackage{algorithmic}
\usepackage{graphicx}
\usepackage{textcomp}
\usepackage{xcolor}
\usepackage{balance}
\usepackage{fancyhdr}
\usepackage{url}
\usepackage{caption}
\usepackage{hyperref}
\usepackage[square,comma,sort&compress,numbers]{natbib} 
\usepackage{threeparttable,multirow,adjustbox,threeparttable}
\usepackage{pbalance} 
\usepackage{hyperref}
\usepackage{adjustbox}
\usepackage{orcidlink}
\usepackage{verbatim}
\hypersetup{
    colorlinks=true,        
    linkcolor=blue,         
    citecolor=blue,         
    filecolor=blue,         
    urlcolor=blue           
}

\usepackage[symbol]{footmisc}

\renewcommand{\arraystretch}{1.3}
\begin{document}
\title{High-Resolution Spatial Transcriptomics from Histology Images using HisToSGE}
\author{
\IEEEauthorblockN{
Zhiceng Shi$^{1,\dag}$, 
Shuailin Xue$^{1,\dag}$,
Fangfang Zhu$^{2,\dag}$ and 
Wenwen Min$^{1,*}\orcidlink{0000-0002-2558-2911}$ }\\
\IEEEauthorblockA{
$^1$School of Information Science and Engineering, Yunnan University, Kunming 650091, Yunnan, China \\
$^2$College of Nursing Health Sciences, Yunnan Open University, 650599, Kunming, China
}
}
\maketitle
\renewcommand{\thefootnote}{\fnsymbol{footnote}} 
\footnotetext[2]{These authors contributed equally to this work.} 
\footnotetext[1]{Corresponding author: minwenwen@ynu.edu.cn}                          
\thispagestyle{fancy}
\begin{abstract}
Spatial transcriptomics (ST) is a groundbreaking genomic technology that enables spatial localization analysis of gene expression within tissue sections. However, it is significantly limited by high costs and sparse spatial resolution. An alternative, more cost-effective strategy is to use deep learning methods to predict high-density gene expression profiles from histological images. However, existing methods struggle to capture rich image features effectively or rely on low-dimensional positional coordinates, making it difficult to accurately predict high-resolution gene expression profiles. To address these limitations, we developed HisToSGE, a method that employs a Pathology Image Large Model (PILM) to extract rich image features from histological images and utilizes a feature learning module to robustly generate high-resolution gene expression profiles. We evaluated HisToSGE on four ST datasets, comparing its performance with five state-of-the-art baseline methods. The results demonstrate that HisToSGE excels in generating high-resolution gene expression profiles and performing downstream tasks such as spatial domain identification. 
All code and public datasets used in this paper are available at \url{https://github.com/wenwenmin/HisToSGE} and \url{https://zenodo.org/records/12792163}.
\end{abstract}


\section{Introduction}
Spatial transcriptomics (ST) is an advanced genomic technology that combines histology and transcriptomics, enabling spatial localization analysis of gene expression on tissue sections \cite{rao2021exploring,chen2022spatiotemporal}. This technology enables researchers to simultaneously observe the spatial heterogeneity of gene expression at the cellular and tissue levels, thereby revealing cell types, functional states, and microenvironment interactions in complex biological processes \cite{spacel,chen2020spatial,cui2023spatiotemporal}. This is of great significance for understanding disease mechanisms, tissue development, and regeneration. \par
Despite the significant advantages of ST technology in biological research, its expensive equipment, reagents, and data analysis costs limit its widespread application \cite{HisToGene}. Furthermore, the sparse distribution of sequencing spots on the sample surface and low spatial resolution result in incomplete or missing gene expression information in certain areas, affecting the comprehensiveness and interpretability of the data \cite{iStar,bayes,xue2024stentrans}. In contrast, whole-slide images \cite{wsi1} stained with hematoxylin and eosin (H\&E) are more convenient and cost-effective, widely used in clinical practice. Predicting ST gene expression profiles using H\&E images has become a common and cost-effective research method \cite{mclSTExp}. Therefore, employing cost-effective methods to address the sparse distribution of sequencing spots and improve spatial resolution in gene expression prediction is crucial for advancing ST technology. \par
Recent approaches have enhanced gene expression prediction by incorporating histological images. For instance, STNet \cite{STnet} employs the DenseNet-121 network to train on image patches corresponding to spots and their gene expression profiles, subsequently predicting gene expression for new spots using these image patches. DeepSpaCE \cite{DeepSpaCE} operates similarly to STNet, but it employs the VGG16 network for gene expression prediction. However, these approaches rely exclusively on histological image information. HistoGene \cite{HisToGene} integrates both histological images and spatial information. It trains a Vision Transformer (ViT) \cite{ViT} on image patches, spatial locations, and gene expression profiles from measured points. It then predicts gene expression for unmeasured points by averaging the gene expression profiles predicted from surrounding image patches. THItoGene \cite{ThItogene} employs dynamic convolution, multi-effect capsule networks, ViT and graph attention networks. This multi-dimensional attention mechanism adaptively captures the intricate relationships among spatial locations, histological images, and gene expression. Although these methods have achieved good performance, the limited number of histological images used for training may result in insufficient extraction of image features. STAGE \cite{STAGE} improves gene expression prediction by combining spatial information and gene expression data. It utilizes a spatially supervised autoencoder generator to produce the two-dimensional or three-dimensional coordinates of unmeasured points, which are then used to generate the corresponding gene expression data. However, low-dimensional positional coordinates may not be sufficient to represent high-dimensional gene expression profiles.\par
\begin{figure*}[htp]%
\centering
\includegraphics[width=1\textwidth]{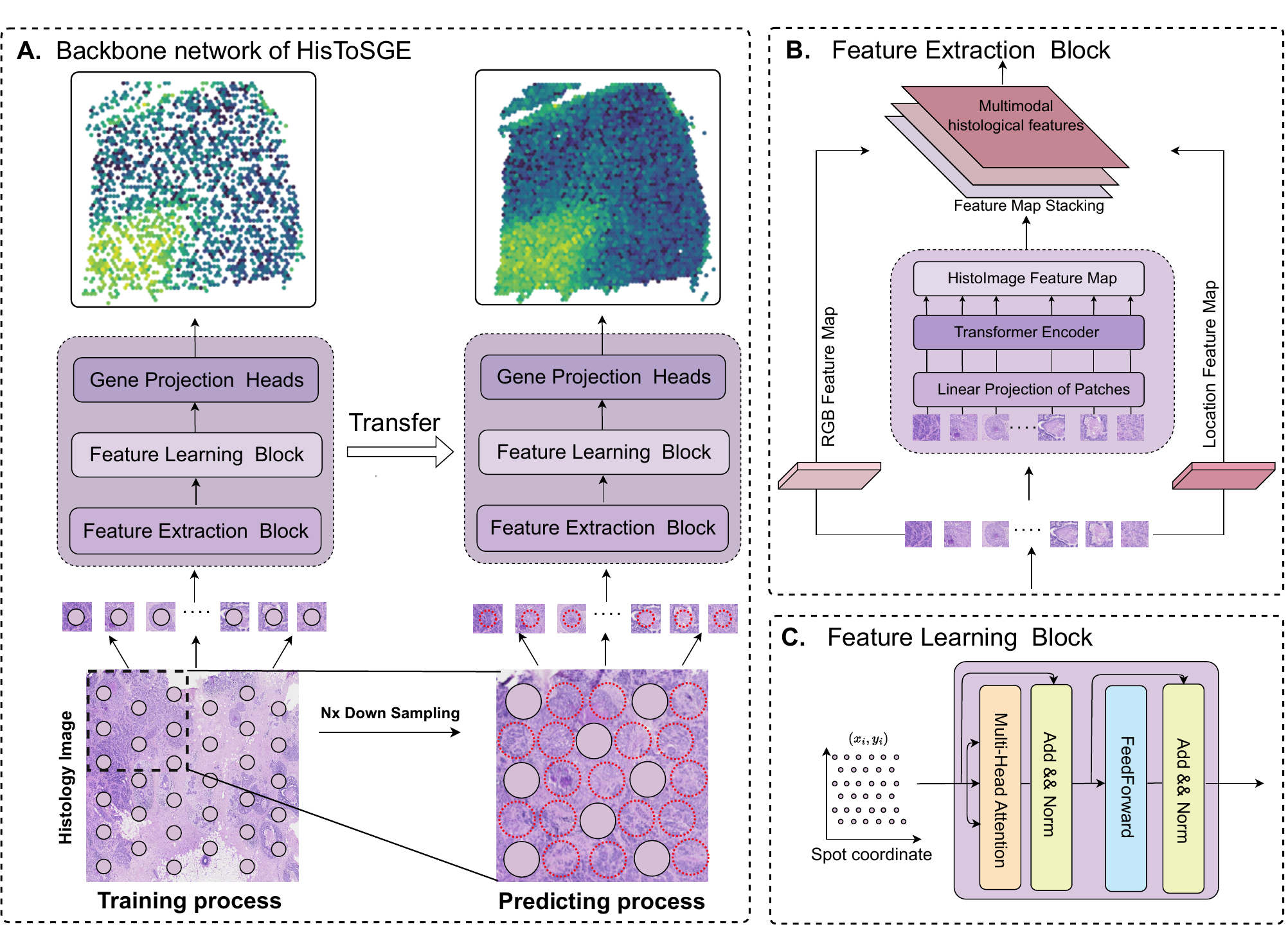}
\caption{The network architecture of our proposed HisToSGE method. 
(\textbf{A}) The backbone network of HisToSGE aims to predict high-resolution gene expression. During training, this network processes histological images through feature extraction and learning modules to associate spatial gene expression patterns at the original resolution. During testing, the histological images are downsampled, and the trained backbone network predicts high-resolution gene expression patterns. 
(\textbf{B}) The feature extraction module generates multimodal feature maps that include RGB, positional, and histological features. 
(\textbf{C}) The feature learning module utilizes a multi-head attention mechanism to integrate spot coordinates and learn features from the multimodal feature maps, thereby enhancing feature representation. }
\label{fig1}
\end{figure*}

To address these issues, we developed HisToSGE, integrates histological image information, spatial information, and gene expression data to robustly generate high-resolution gene expression profiles in ST. HisToSGE comprises two main modules: the feature extraction module and the feature learning module (\autoref{fig1}). The feature extraction module, utilizing the UNI \cite{UNI} model trained on one hundred million histological images, generates multimodal feature maps that include RGB, positional, and histological features. The feature learning module employs a multi-head attention mechanism to integrate spot coordinates and learn features from these multimodal maps, thereby enhancing feature representation. We evaluated HisToSGE using ST four datasets and compared its performance with five existing methods. Our results demonstrate that HisToSGE can accurately generate high-resolution gene expression profiles, enhance gene expression patterns, and preserve the original gene expression spatial structure. \par
The main contributions of our proposed method are:
\begin{itemize}
  \item  We developed HisToSGE, a deep learning-based model that predicts high-resolution spatial gene expression profiles from histological images. HisToSGE employs a large histological image model to extract rich image features and uses a feature learning module to integrate the spatial positions of spots, enhancing feature representation.
  
  \item Our method was compared with other approaches on multiple real ST datasets. The results demonstrate that our method improves the average Pearson Correlation Coefficient (PCC) by 9\% to 32\% in generating high-density gene expression profiles compared to state-of-the-art methods. Additionally, Our approach not only enhances the original gene expression patterns but also effectively preserves the original spatial structure.

  \item Compared to other image-to-gene expression methods, HisToSGE can generate spatial gene expression profiles at any desired resolution.
\end{itemize}
\section{PROPOSED METHODS}\label{MATERIALS AND MRTHODS}
\subsection{Overview of the proposed HisToSGE}
We propose HisToSGE to enhance the resolution of spatial gene expression using histological images. During the training phase, we divide the histological images into patches based on the original resolution coordinates of the spots. HisToSGE starts with the Feature Extraction Module to obtain multimodal image feature maps. These feature maps are then processed by the Feature Learning Module. Finally, the Gene Projection Heads project the processed image features into the gene expression dimension. In the testing phase, we downsample the spot coordinates N-fold to achieve high-resolution spot positions, which are then divided into image patches. Using the trained HisToSGE model, we generate high-resolution gene expression data from these patches (\autoref{fig1}).

\subsection{HisToSGE model}
\subsubsection{Feature Extraction Block of HisSGE}
The Feature Extraction Block utilizes the UNI model \cite{UNI} trained on a large set of histological images to generate multimodal feature maps that include RGB, positional, and histological features.\par
We segment 50 $\times$ 50 pixel image patches from histological images based on the positions of spots as input. Each patch can be represented as $\text{Patch}_{i}\in \mathbb{R}^{50}\times \mathbb{R}^{50}\times \mathbb{R}^{3}$, where $i\in [1, \text{spot}\_\text{num}]$.
\begin{equation}
    \text{Z}_{i}=\text{UNI}(\text{Patch}_{i}), \text{Z}_{i}\in \mathbb{R}^{1}\times \mathbb{R}^{1024}
\end{equation}
where $\text{Z}_{i}$ is the histological feature map.\par
Since the spot coordinates correspond directly to the pixel coordinates in the image, we use the corresponding image pixel coordinates as the location feature map $\text{L}_{i} \in \mathbb{R} ^{1\times2}$.\par
Additionally, we obtain the RGB feature map $\text{T}_{i} \in \mathbb{R} ^{1\times3}$ by resizing the entire histological image to the desired size $50 \times 50$ using average pooling. From this point, we have obtained histological feature map $\text{Z}_{i}$, location feature map $\text{L}_{i}$, and RGB feature map $\text{T}_{i}$. By stacking these feature maps along the channel dimension, we obtain the multimodal feature map:
\begin{equation}
    \text{M}_{i}=\text{concat}\left (\text{Z}_{i},\text{L}_{i},\text{T}_{i} \right )  \in \mathbb{R} ^{1}\times \mathbb{R} ^{1024+2+3}
\end{equation}
where $\text{M}_{i}$ represents the multimodal feature map.

\subsubsection{The Feature Learning of HisSGE}
The Feature Learning Block uses a multi-head attention mechanism to integrate spot coordinates and learn features from the multimodal feature maps. This block includes layers for multi-head attention to enhance feature representation.\par
Multi-head attention is an extension of the attention mechanism, enhancing the model's ability to capture complex patterns and global information in input sequences by simultaneously learning multiple independent sets of attention weights as follows:
\begin{equation}\label{equ:multihead_attention}
	\begin{aligned}
		\text{MHSA}(Q, K, V) = [head_1, \ldots, head_n] W_0
	\end{aligned}
\end{equation}
where $W_{0}$ represents the weight matrix used for aggregating the attention heads, while $n$ denotes the number of heads. Additionally, $Q$, $K$, and $V$ correspond to Query, Key, and Value, respectively.
The attention mechanism is defined as follows:
\begin{equation}\label{equ:attention_mechanism}
	\begin{aligned}
		\text{head}_i = \text{Attention}(QW_{i}^{Q} , KW_{i}^{K}, VW_{i}^{V})
	\end{aligned}
\end{equation}
\begin{equation}\label{equ:attention_mechanism}
	\begin{aligned}
		\text{Attention}(Q, K, V) = \text{softmax}(\frac{QK^{T}}{\sqrt{d_{k} }})V
	\end{aligned}
\end{equation}
where $W_{i}^{Q}$, $W_{i}^{K}$ and $W_{i}^{V}$ are weight matrices. The term $(\frac{\text{QK}^{\text{T}}}{\sqrt{\text{d}_{\text{k}}}})$ is called Attention Map, whose shape is $N\times N$. The term $V$ is the value of the self-attention mechanism, where $V = Q = K$.\par
For the multimodal feature map $\text{M}_{i}$, we further refine the feature representations using the multi-head attention mechanism and integrate spot location information with learnable positional encodings:
\begin{equation}
    \text{H}_{i}=\text{MHSA}(\text{M}_{i} + \text{PE}_{i}), \text{H}_{i}\in \mathbb{R}^{1}\times \mathbb{R}^{1024+2+3}
\end{equation}
where $\text{PE}_{i}$ represents the positional encoding of $\text{spot}_{i}$.
\subsubsection{The Gene Projection Heads of HisSGE}
The Gene Projection Heads take the learned features from the Feature Learning Block and project them into the gene expression dimension.
\begin{equation}
    \begin{aligned}
         \text{X}_{\text{pred}} = \text{MLP}(\text{H}_{i}),\text{X}_{\text{pred}}\in \mathbb{R}^{1}\times \mathbb{R}^{gene\_dim}
    \end{aligned}
\end{equation}
where $\text{X}_{\text{pred}}$ represents the predicted gene expression.
\subsubsection{The Loss function of HisSGE}
The loss function of HisToSGE is designed to minimize the difference between the predicted and actual gene expression data. It ensures the model accurately generates high-resolution gene expression profiles from the input histological images. We apply the mean square errors loss as follows:
\begin{equation}
    \begin{aligned}
        \text{Loss}=\sum \left \| \text{X}_{\text{observed}} -  \text{X}_{\text{pred}} \right \|^{2} _{2}
    \end{aligned}
\end{equation}
where $\text{X}_{\text{observed}}$ and $\text{X}_{\text{pred}}$ are the observed and predicted gene expression, respectively.

\subsection{Evaluation metrics}\label{Evaluation metrics}
We use Pearson Correlation Coefficient (PCC), Mean Squared Error (MSE), and Mean Absolute Error (MAE) to evaluate the proposed method against baselines. 
\begin{equation}\label{equ:pcc}
    \begin{aligned}
         \text{PCC}=\frac{\text{Cov}(\text{X}_{\text{observed}},~\text{X}_{\text{pred}})}{\text{Var}(\text{X}_{\text{observed}}) \times \text{Var}(\text{X}_{\text{pred}})}
    \end{aligned}
\end{equation}
where $\text{Cov()}$ is the covariance, and $\text{Var()}$ is the variance.
$\text{X}_\text{observed}$ and $\text{X}_\text{pred}$ are the observed and predicted gene expression, respectively.
\begin{equation}
    \begin{aligned}
        \text{MSE}=\frac{1}{N} \sum_{i=1}^{N} (\text{X}_{\text{observed}}-\text{X}_{\text{pred}})^{2}
    \end{aligned}
\end{equation}
\begin{equation}
    \begin{aligned}
        \text{MAE}=\frac{1}{N} \sum_{i=1}^{N} \left | \text{X}_{\text{observed}}-\text{X}_{\text{pred}}\right |
    \end{aligned}
\end{equation}
In the assessment of spatial clustering performance, we employ the Adjusted Rand Index (ARI) to measure the correlation between the clustering outcomes and the actual pathological annotation regions. The ARI can be mathematically expressed as follows:
\begin{equation}
\text{ARI}= \frac{ {\textstyle \sum_{ij}\binom{n_{ij}}{2}}-\frac{\left [ {\textstyle \sum_{i}\binom{a_{i}}{2}\sum_{j}\binom{b_{j}}{2}}\right] }{\binom{n}{2}}}{\frac{1}{2} \left  [\sum_{i}\binom{a_{i}}{2}+\sum_{j}\binom{b_{j}}{2}\right ]- \frac{\left [ {\textstyle \sum_{i}\binom{a_{i}}{2}\sum_{j}\binom{b_{j}}{2}}\right] }{\binom{n}{2}}}
\end{equation}
where $a_{i}$ and $b_{j}$ are the number of samples appearing in the $i-th$ predicted cluster and the $j-th$ true cluster, respectively. $n_{ij}$ means the number of overlaps between the $i-th$ predicted cluster and the
$j-th$ true cluster.\par

\begin{table}[htp]
\caption{Summary of the ST datasets used in this study.}
\label{tab1}
\begin{adjustbox}{width=0.49\textwidth}
\fontsize{9}{10}\selectfont    
\setlength{\arrayrulewidth}{0.05mm} 
\renewcommand{\arraystretch}{1.2}
\begin{tabular}{l|c|c|c|c}
\hline
\textbf{Datasets}   & \textbf{Spots} & \textbf{Genes} & \textbf{Dropout rate} & \textbf{Ref.}\\ \hline
              
DLPFC (slice 151507)          & 4226  & 33538 & 0.96 & \cite{DLPFC}    \\
DLPFC (slice 151508)          & 4384  & 33538 & 0.96 & \cite{DLPFC}    \\
DLPFC (slice 151509)          & 4789  & 33538 & 0.96 & \cite{DLPFC}    \\
DLPFC (slice 151510)          & 4634  & 33538 & 0.96 & \cite{DLPFC}    \\
DLPFC (slice 151669)          & 3661  & 33538 & 0.96 & \cite{DLPFC}    \\
DLPFC (slice 151670)          & 3498  & 33538 & 0.95 & \cite{DLPFC}    \\
DLPFC (slice 151671)          & 4110  & 33538 & 0.94 & \cite{DLPFC}    \\
DLPFC (slice 151672)          & 4015  & 33538 & 0.95 & \cite{DLPFC}    \\
DLPFC (slice 151673)          & 3639  & 33538 & 0.93 & \cite{DLPFC}    \\
DLPFC (slice 151674)          & 3673  & 33538 & 0.92 & \cite{DLPFC}    \\
DLPFC (slice 151675)          & 3592  & 33538 & 0.95 & \cite{DLPFC}    \\
DLPFC (slice 151676)          & 3460  & 33538 & 0.94 & \cite{DLPFC}    \\ \hline
Mouse Brain & 4992  & 21949 & 0.93 & \cite{MouseBrain}    \\
Human Breast Cancer1 (BC1) & 3813  & 22968 & 0.95  & \cite{10x}   \\
Human Breast Cancer2 (BC2)    & 2518  & 17651 & 0.95  & \cite{10x}   \\ \hline
\end{tabular}
\end{adjustbox}
\end{table}

\begin{table*}[htp]
\centering
\caption{Comparison with baseline methods on the four ST datasets (See \autoref{tab1}). We use three evaluation indicators to comprehensively measure the recovery performance of gene expression using these methods. The proposed HisToSGE achieves state-of-the-art performance.}
\label{tab2}
\begin{adjustbox}{width=1\textwidth}
\fontsize{9}{10}\selectfont          
\setlength{\arrayrulewidth}{0.05mm}  
\begin{tabular}{l|cccccccccccc|c|c|c}
\hline
\multirow{2}{*}{\textbf{PCC}\hspace{0em} $\uparrow$}    & \multicolumn{12}{c|}{SectionID of DLPFC datasets}  & \multirow{2}{*}{Mouse Brain} & \multirow{2}{*}{ BC1 } & \multirow{2}{*}{ BC2 } \\
                        & 151507          & 151508          & 151509          & 151510          & 151669          & 151670          & 151671          & 151672          & 151673          & 151674          & 151675          & 151676          &                              &                                &                       \\ \hline
STnet \cite{STnet}                   & 0.1221          & 0.1092          & 0.1301          & 0.1143          & 0.1042          & 0.1012          & 0.1009          & 0.1113          & 0.1075          & 0.1051          & 0.1091          & 0.1072          & 0.2615                       & 0.2012                         & 0.0976                \\
DeepSpace \cite{DeepSpaCE}               & 0.2080          & 0.2291          & 0.2151          & 0.2126          & 0.2094          & 0.2319          & 0.2155          & 0.2665          & 0.2523          & 0.2717          & 0.2616          & 0.2087          & 0.3082                       & 0.2685                         & 0.2455                \\
HistoGene \cite{HisToGene}               & 0.3189          & 0.2254          & 0.3726          & 0.3926          & 0.3676          & 0.1178          & 0.4425          & 0.2643          & 0.3003          & 0.4564          & 0.2067          & 0.4014          & 0.3384                       & 0.2463                         & 0.1186                \\
THItoGene \cite{ThItogene}               & 0.3509          & 0.3233          & 0.4142          & 0.2619          & 0.4129          & 0.4231          & 0.2155          & 0.3456          & 0.4054          & 0.4416          & 0.1769          & 0.4512          & 0.3875                       & 0.3051                         & 0.3446                \\
STAGE \cite{STAGE}                   & 0.3944          & 0.4147          & 0.4332          & 0.4233          & 0.4437          & 0.4362          & 0.2638          & 0.4455          & 0.5222          & 0.5033          & 0.4911          & 0.4869          & 0.6598                       & 0.4765                         & 0.6034                \\
\textbf{HisToSGE (ours)} & \textbf{0.6511} & \textbf{0.6556} & \textbf{0.6545} & \textbf{0.6391} & \textbf{0.6474} & \textbf{0.5018} & \textbf{0.6501} & \textbf{0.6096} & \textbf{0.6725} & \textbf{0.6773} & \textbf{0.6650} & \textbf{0.6705} & \textbf{0.7268}              & \textbf{0.6937}                & \textbf{0.7117}       \\ \hline
\hline
\multirow{2}{*}{\textbf{MSE}\hspace{0em} $\downarrow$}    & \multicolumn{12}{c|}{SectionID of DLPFC datasets}                                                                                                                                                                                  & \multirow{2}{*}{Mouse Brain} & \multirow{2}{*}{ BC1 } & \multirow{2}{*}{ BC2 } \\
                        & 151507          & 151508          & 151509          & 151510          & 151669          & 151670          & 151671          & 151672          & 151673          & 151674          & 151675          & 151676          &                              &                                &                       \\ \hline
STnet \cite{STnet}                   & 0.2113          & 0.2135          & 0.2321          & 0.2089          & 0.2132          & 0.2156          & 0.1824          & 0.1964          & 0.2204          & 0.2168          & 0.2455          & 0.1897          & 0.1939                       & 0.1701                         & 0.2837                \\
DeepSpace \cite{DeepSpaCE}                & 0.2046          & 0.2076          & 0.2204          & 0.2034          & 0.2048          & 0.2019          & 0.1581          & 0.1859          & 0.2196          & 0.1946          & 0.2348          & 0.1788          & 0.1813                       & 0.1638                         & 0.2608                \\
HistoGene \cite{HisToGene}               & 0.1725          & 0.1763          & 0.1936          & 0.1731          & 0.1332          & 0.1729          & 0.1358          & 0.1649          & 0.1791          & 0.1542          & 0.2059          & 0.1627          & 0.1657                       & 0.1521                         & 0.2375                \\
THItoGene \cite{ThItogene}               & 0.1572          & 0.1702          & \textbf{0.1447} & 0.1455          & 0.1301          & 0.1589          & 0.1351          & 0.1601          & 0.1473          & 0.1411          & 0.1924          & 0.1635          & 0.1536                       & 0.1353                         & 0.2647                \\
STAGE\cite{STAGE}                   & 0.1461          & 0.1504          & 0.1711          & 0.1633          & 0.1435          & 0.1562          & 0.1325          & 0.1427          & 0.1647          & 0.1444          & 0.1832          & 0.1702          & 0.1189                       & 0.1036                         & 0.1650                \\
\textbf{HisToSGE (ours)} & \textbf{0.1374} & \textbf{0.1483} & 0.1566          & \textbf{0.1423} & \textbf{0.1285} & \textbf{0.1556} & \textbf{0.1147} & \textbf{0.1324} & \textbf{0.1416} & \textbf{0.1307} & \textbf{0.1638} & \textbf{0.1470} & \textbf{0.1064}              & \textbf{0.0917}                & \textbf{0.1578}       \\ \hline
\hline
\multirow{2}{*}{\textbf{MAE}\hspace{0em} $\downarrow$}    & \multicolumn{12}{c|}{SectionID of DLPFC datasets}                                                                                                                                                                                  & \multirow{2}{*}{Mouse Brain} & \multirow{2}{*}{BC1} & \multirow{2}{*}{BC2} \\
                        & 151507          & 151508          & 151509          & 151510          & 151669          & 151670          & 151671          & 151672          & 151673          & 151674          & 151675          & 151676          &                              &                                &                       \\ \hline 
STnet \cite{STnet}                   & 0.2362          & 0.2316          & 0.2321          & 0.2251          & 0.2366          & 0.2216          & 0.2049          & 0.2171          & 0.2585          & 0.2444          & 0.2509          & 0.2503          & 0.2707                       & 0.2311                         & 0.3966                \\
DeepSpace \cite{DeepSpaCE}                & 0.2116          & 0.2137          & 0.2204          & 0.2127          & 0.2241          & 0.2246          & 0.1879          & 0.2088          & 0.2527          & 0.2401          & 0.2568          & 0.2435          & 0.2174                       & 0.2259                         & 0.3581                \\
HistoGene \cite{HisToGene}               & 0.1730          & 0.1808          & 0.1981          & 0.1801          & 0.1768          & 0.1896          & 0.1533          & 0.1807          & 0.2216          & 0.1835          & 0.2311          & 0.2001          & 0.2540                       & 0.2289                         & 0.3786                \\
THItoGene \cite{ThItogene}               & 0.1346          & 0.1427          & 0.1471          & 0.1483          & 0.1481          & 0.1464          & 0.1491          & 0.1675          & 0.1796          & 0.1699          & 0.2033          & 0.1959          & 0.2411                       & 0.2148                         & 0.2148                \\
STAGE \cite{STAGE}                   & 0.1283          & 0.1268          & 0.1351          & 0.1271          & 0.1342          & 0.1381          & 0.1296          & 0.1283          & 0.1643          & 0.1661          & 0.1633          & 0.1631          & 0.1563                       & 0.1382                         & 0.3598                \\
\textbf{HisToSGE (ours)} & \textbf{0.0907} & \textbf{0.0928} & \textbf{0.1013} & \textbf{0.0926} & \textbf{0.1003} & \textbf{0.1169} & \textbf{0.1006} & \textbf{0.1010} & \textbf{0.1223} & \textbf{0.1234} & \textbf{0.1217} & \textbf{0.1168} & \textbf{0.1347}              & \textbf{0.1209}                & \textbf{0.1952}       \\ \hline
\end{tabular}

\end{adjustbox}
\end{table*}

\begin{table*}[htp]
\caption{Ablation studies on HisToSGE with different backbone networks. In STAGE\_Plus, the input of STAGE is replaced with image features extracted by a large-scale histological image model, while all other components remain unchanged.}
\label{tab3}
\begin{adjustbox}{width=1\textwidth}
\fontsize{9}{10}\selectfont            
\setlength{\arrayrulewidth}{0.05mm}    
\begin{tabular}{l|cccccccccccc|c|c|c}
\hline
\multirow{2}{*}{\textbf{PCC}\hspace{0em} $\uparrow$} & \multicolumn{12}{c|}{SectionID of DLPFC datasets}                                                                                                                                                                                  & \multirow{2}{*}{Mouse Brain} & \multirow{2}{*}{BC1} & \multirow{2}{*}{BC2} \\
                     & 151507          & 151508          & 151509          & 151510          & 151669          & 151670          & 151671          & 151672          & 151673          & 151674          & 151675          & 151676          &                              &                                &                       \\ \hline 
STAGE\_Plus           & 0.4038          & 0.3547          & 0.4781          & 0.4571          & 0.4502          & 0.4511          & 0.4187          & 0.4649          & 0.5581          & 0.5587          & 0.5146          & 0.5055          & 0.6453                       & 0.5765                         & 0.6353                \\
w/ FeedForward          & 0.5970          & 0.6049          & 0.5152          & 0.6089          & 0.6025          & 0.4160          & 0.3304          & 0.5295          & 0.6647          & 0.6503          & 0.5909          & 0.6462          & 0.7199                       & 0.6423                         & 0.6546                \\
w/ GAT                  & 0.6274          & 0.6253          & 0.6055          & 0.5766          & 0.6179          & 0.4895          & 0.6082          & 0.5869          & 0.6276          & 0.6307          & 0.6272          & 0.6390          & 0.6984                       & 0.6802                         & 0.6989                \\
\textbf{w/ Transformer}  & \textbf{0.6589} & \textbf{0.6556} & \textbf{0.6557} & \textbf{0.6391} & \textbf{0.6474} & \textbf{0.5018} & \textbf{0.6501} & \textbf{0.6096} & \textbf{0.6725} & \textbf{0.6773} & \textbf{0.6650} & \textbf{0.6705} & \textbf{0.7268}              & \textbf{0.6937}                & \textbf{0.7117}       \\ \hline \hline
\multirow{2}{*}{\textbf{MSE}\hspace{0em} $\downarrow$} & \multicolumn{12}{c|}{SectionID of DLPFC datasets}                                                                                                                                                                                  & \multirow{2}{*}{Mouse Brain} & \multirow{2}{*}{BC1} & \multirow{2}{*}{BC2} \\
                     & 151507          & 151508          & 151509          & 151510          & 151669          & 151670          & 151671          & 151672          & 151673          & 151674          & 151675          & 151676          &                              &                                &                       \\ \hline 
STAGE\_Plus           & 0.1639          & 0.1772          & 0.1796          & 0.1703          & 0.1517          & 0.1646          & 0.1388          & 0.1491          & 0.1725          & 0.1522          & 0.1944          & 0.1807          & 0.1235                       & 0.1023                         & 0.1623                \\
w/ FeedForward          & 0.1439          & 0.1544          & 0.1614          & 0.1533          & 0.1357          & 0.1451          & 0.1437          & 0.1347          & 0.1449          & 0.1291          & 0.1586          & 0.1546          & 0.1146                       & 0.1132                         & 0.1615                \\
w/ GAT                  & 0.1423          & 0.1595          & 0.1566          & 0.1453          & 0.1285          & 0.1381          & 0.1226          & 0.1345          & 0.1502          & 0.1378          & 0.1675          & 0.1580          & 0.1221                       & 0.1048                         & 0.1589                \\
\textbf{w/ Transformer}  & \textbf{0.1374} & \textbf{0.1483} & \textbf{0.1533} & \textbf{0.1423} & \textbf{0.1261} & \textbf{0.1556} & \textbf{0.1147} & \textbf{0.1324} & \textbf{0.1416} & \textbf{0.1307} & \textbf{0.1638} & \textbf{0.1470} & \textbf{0.1064}              & \textbf{0.0917}                & \textbf{0.1578}       \\ \hline \hline
\multirow{2}{*}{\textbf{MAE}\hspace{0em} $\downarrow$} & \multicolumn{12}{c|}{SectionID of DLPFC datasets}                                                                                                                                                                                  & \multirow{2}{*}{Mouse Brain} & \multirow{2}{*}{BC1} & \multirow{2}{*}{BC2} \\
                     & 151507          & 151508          & 151509          & 151510          & 151669          & 151670          & 151671          & 151672          & 151673          & 151674          & 151675          & 151676          &                              &                                &                       \\ \hline
STAGE\_Plus           & 0.1348          & 0.1341          & 0.1446          & 0.1368          & 0.1446          & 0.1496          & 0.1294          & 0.1385          & 0.1781          & 0.1792          & 0.1783          & 0.1774          & 0.1564                       & 0.1367                         & 0.0.3342              \\
w/ FeedForward          & 0.1009          & 0.1003          & 0.1068          & 0.1002          & 0.1047          & 0.1088          & 0.1267          & 0.1289          & 0.1285          & 0.1302          & 0.1273          & 0.1248          & 0.1469                       & 0.1289                         & 0.2416                \\
w/ GAT                  & 0.0942          & 0.0952          & 0.1036          & 0.0975          & 0.1043          & \textbf{0.1059} & 0.0924          & 0.1267          & 0.1246          & 0.1263          & 0.1232          & 0.1228          & 0.1412                       & 0.1245                         & 0.2287                \\
\textbf{w/ Transformer}  & \textbf{0.0907} & \textbf{0.0916} & \textbf{0.1013} & \textbf{0.0926} & \textbf{0.1003} & 0.1169          & \textbf{0.0891} & \textbf{0.1012} & \textbf{0.1223} & \textbf{0.1234} & \textbf{0.1217} & \textbf{0.1168} & \textbf{0.1347}              & \textbf{0.1209}                & \textbf{0.1952}  \\ \hline    
\end{tabular}
\end{adjustbox}
\end{table*}

\begin{figure*}[htp]%
\centering
\includegraphics[width=1\textwidth]{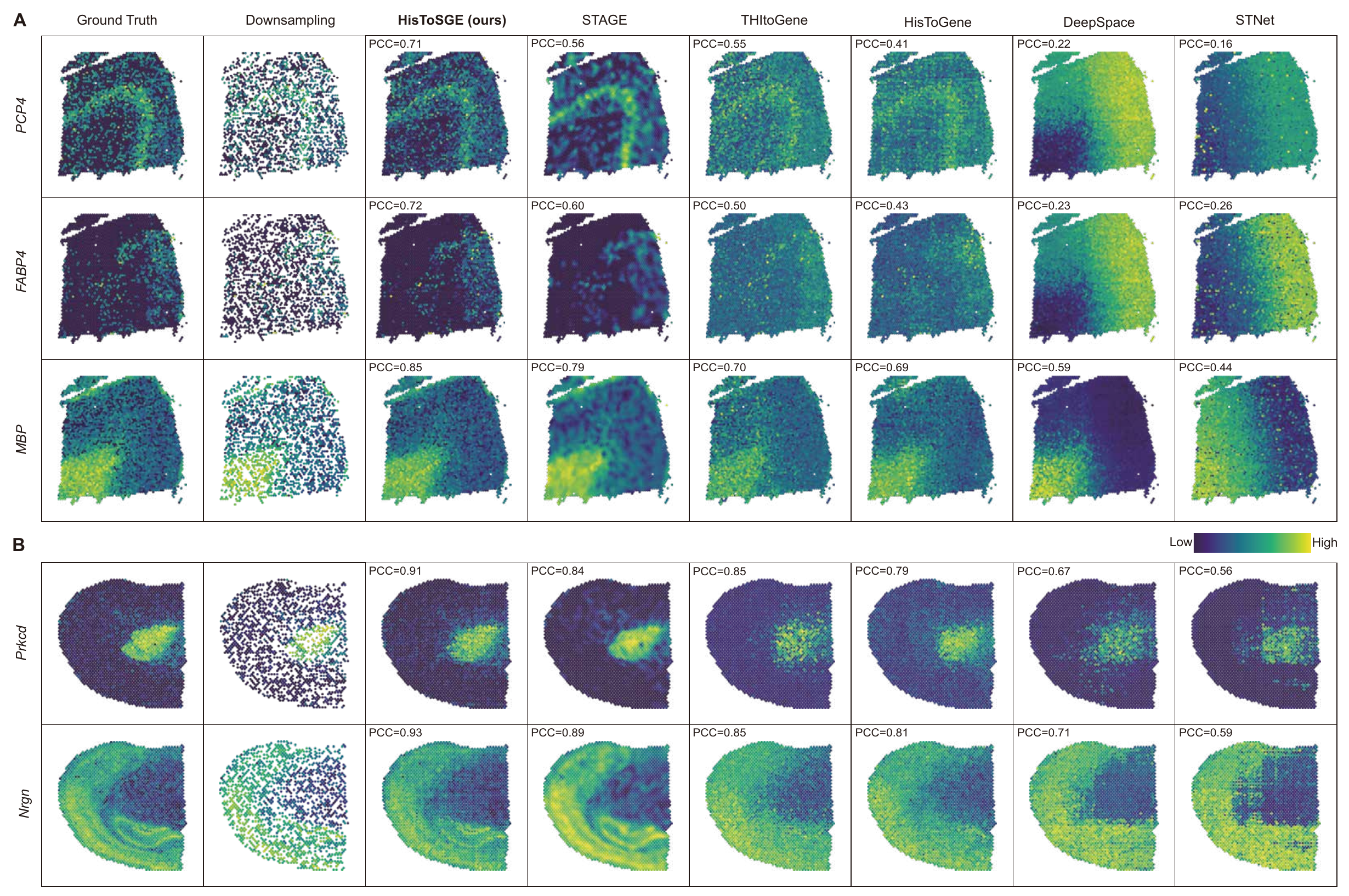}
\caption{HisToSGE excels in generating gene expression profiles in tissues. (\textbf{A}) For visualizing the marker genes \textit{PCP4}, \textit{FABP4}, and \textit{MBP} in downsampled and generated data, methods like HisToSGE, STAGE, THItoGene, HisToGene, DeepSpaCE, and STNet were used on slice 151676 from the DLPFC dataset. (\textbf{B}) Similarly, for visualizing the marker genes \textit{Prkcd} and \textit{Mrgn} in downsampled and generated data, the same methods were applied to the Mouse Brain dataset.}
\label{fig2}
\end{figure*}

\begin{figure*}[htp]%
\centering
\includegraphics[width=1\textwidth]{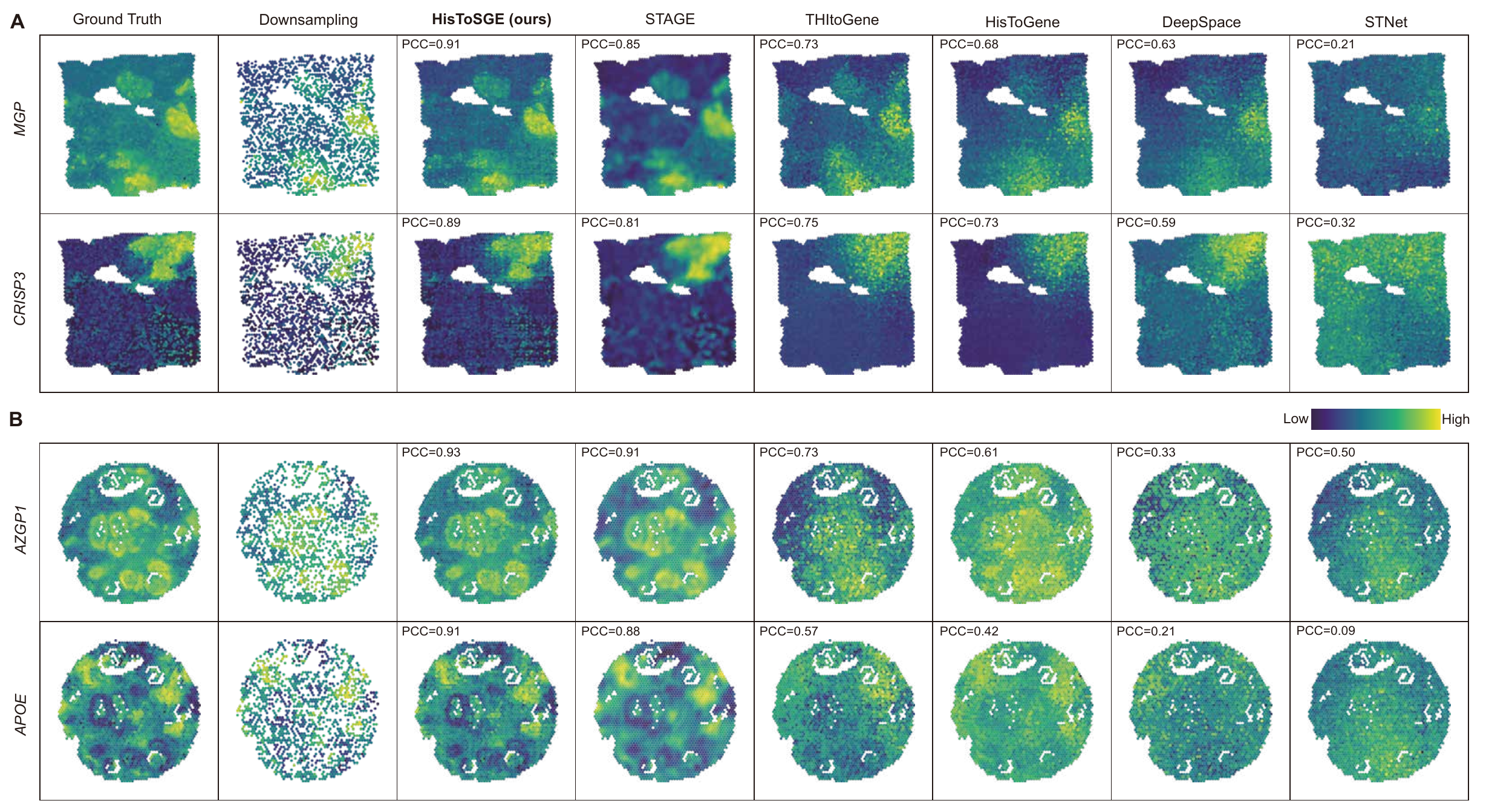}
\caption{(\textbf{A}) For visualizing the marker genes \textit{MGP} and \textit{CRISP3} in downsampled and generated data, methods like HisToSGE, STAGE, THItoGene, HisToGene, DeepSpaCE, and STNet were used on the BC1 dataset. (\textbf{B}) Similarly, for visualizing the marker genes \textit{AZGP1} and \textit{APOE} in downsampled and generated data, the same methods were applied to the BC2 dataset.}
\label{fig3}
\end{figure*}

\begin{figure}[htp]%
\centering
\includegraphics[width=0.48\textwidth]{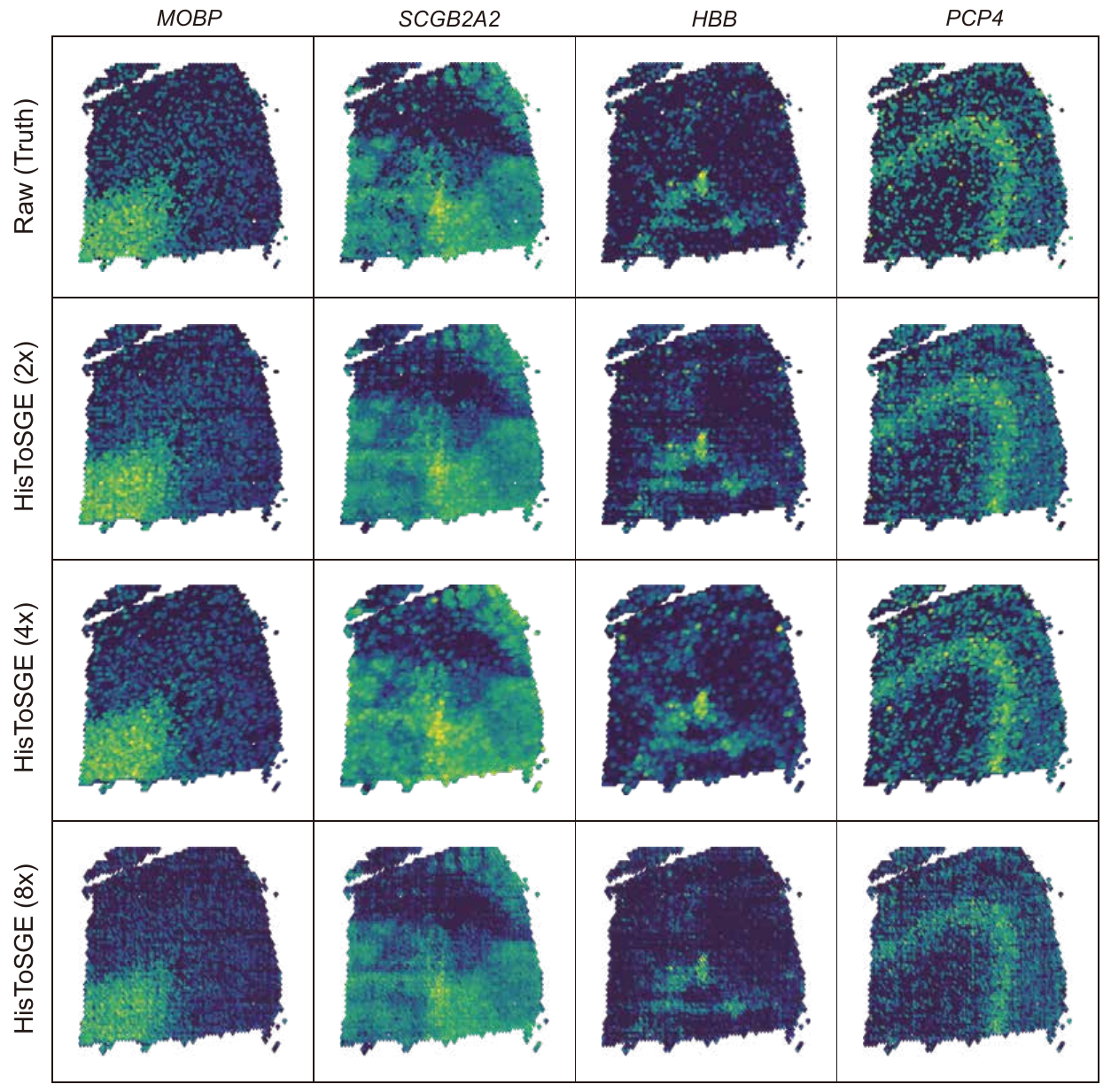}
\caption{HisToSGE enhances gene expression patterns in the DLPFC dataset. Spatial visualization of marker genes \textit{MOBP}, \textit{SCGB2A2}, \textit{HBB} and \textit{PCP4} for the raw and generated data, respectively}
\label{fig4}
\end{figure}
\begin{figure*}[htp]%
\centering
\centering
\includegraphics[width=1\textwidth]{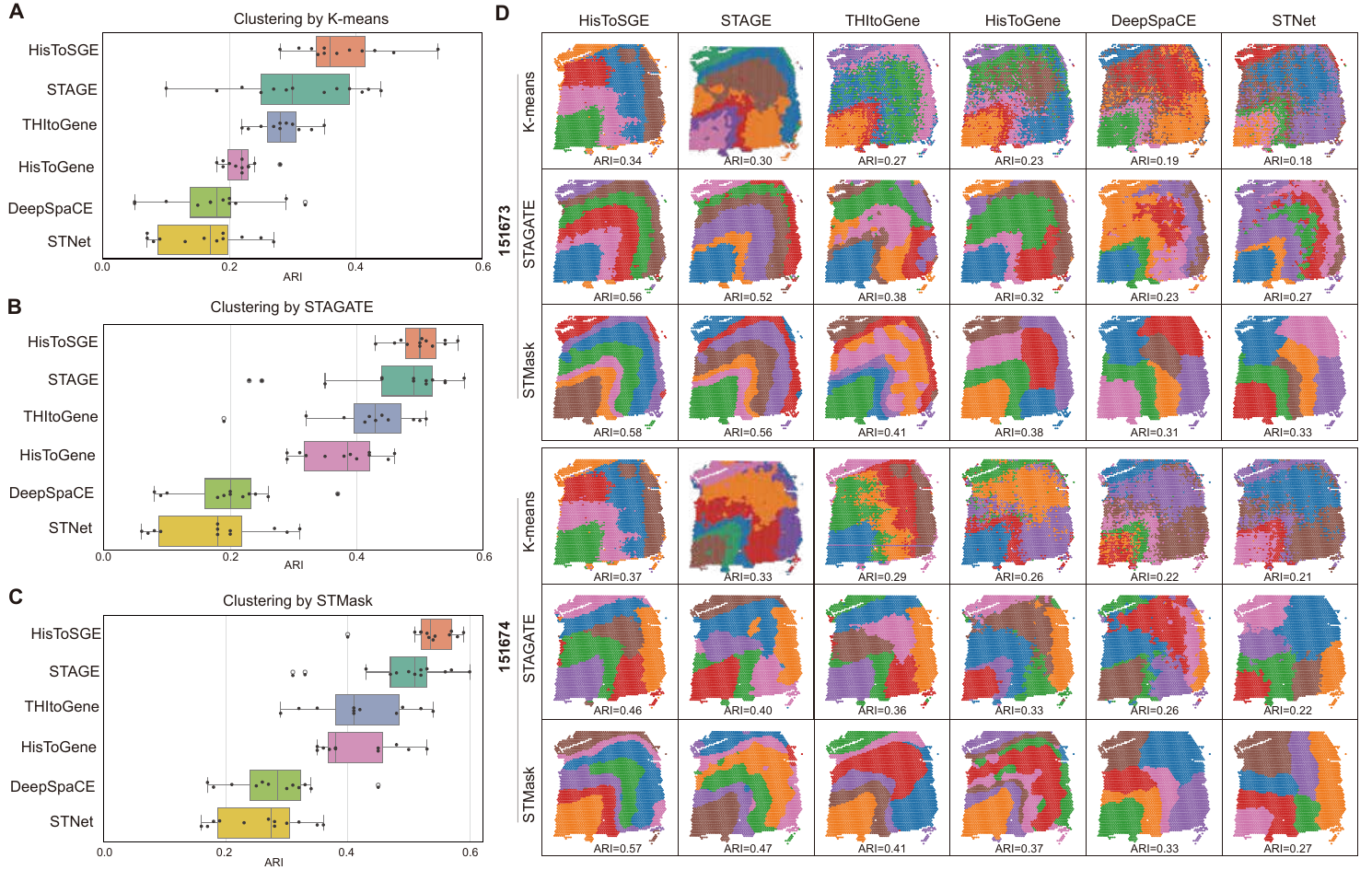}
\caption{Accuracy of spatial domains identified from raw data and data generated by HisToSGE, STAGE \cite{STAGE}, THItoGene \cite{ThItogene}, HisToGene \cite{HisToGene}, DeepSpaCE \cite{DeepSpaCE}, and ST-Net \cite{STnet} on the DLPFC dataset. Performance comparison of identification accuracy by HisToSGE and other methods when employing K-means (\textbf{A}) , STAGATE \cite{STGATE} (\textbf{B}) and STmask \cite{STMask} (\textbf{C}). (\textbf{D}) Spatial visualization of identified spatial domains by HisToSGE and other methods for slices 151673 and 151674, respectively.}
\label{fig5}
\end{figure*}

\section{EXPERIMENTAL RESULTS}\label{result}
\subsection{Dataset and pre-processing}\label{Dataset and pre-processing}
Four publicly available ST datasets were used in this study (\autoref{tab1}). 
\begin{itemize}
  \item \textbf{DLPFC dataset} \cite{DLPFC} consists of 12 sections of the dorsolateral prefrontal cortex (DLPFC) sampled from three individuals. The number of spots for each section ranges from 3498 to 4789. The original authors have manually annotated the areas of the DLPFC layers and white matter. The datasets are available in the \href{http://spatial.libd.org/spatialLIBD}{spatialLIBD} package.

  \item \textbf{MouseBrain dataset} \cite{MouseBrain} includes a coronal brain section sample from an adult mouse, with 2903 sampled spots. The datasets are available in \href{https://cf.10xgenomics.com/samples/spatial-exp/1.1.0/V1_Adult_Mouse_Brain/V1_Adult_Mouse_Brain_web_summary.html}{10x Genomics Website}.
  
  \item \textbf{Human Breast Cancer1 (BC1) dataset} \cite{10x} includes a fresh frozen invasive ductal carcinoma breast tissue section sample, with 3813 sampled spots. The datasets are available in \href{https://www.10xgenomics.com/datasets/human-breast-cancer-block-a-section-1-1-standard-1-0-0}{10x Genomics Website}.
  
  \item \textbf{Human Breast Cancer2 (BC2) dataset} \cite{10x} includes a formalin-fixed invasive breast carcinoma tissue section sample, with 2518 sampled spots. The datasets are available in \href{https://www.10xgenomics.com/datasets/human-breast-cancer-ductal-carcinoma-in-situ-invasive-carcinoma-ffpe-1-standard-1-3-0}{10x Genomics Website}.
\end{itemize}

\textbf{Image data preprocessing.}
For H\&E images, we partitioned a $W \times H$ pixel region around each sequencing spot based on its positional coordinates. Both $W$ and $H$ are set to 50.\par
\textbf{Gene expression data preprocessing.}
The raw gene expression counts were normalized and log-transformed. The HisToSGE model uses the top 1,000 highly variable genes from the normalized data as input.\par
\textbf{Construction of unmeasured spots.} To maximize coverage of tissue sections, we translate the measurement points in $N -1$ specific directions by $N -1$ specific distances according to the sampling multiple $N$, thereby constructing unmeasured points. Given that the translation direction and distance for each measurement point are identical, we utilize polar coordinates to describe the relationship between the unmeasured points and the measurement points. Let $R$ be the Euclidean distance between adjacent measured spots. We denote $r$ and $\theta$ as the polar radius and polar angle of the polar coordinate system. Specifically, for an 8-fold downsampling, we translated the measured spots to ($r$,$\theta$) = ($\frac{R}{2},0$), ($\frac{R}{2\sqrt{2}},\frac{\pi}{4}$), ($\frac{R}{2\sqrt{2}},\frac{3\pi}{4}$), ($\frac{R}{2\sqrt{2}},\frac{7\pi}{4}$), ($\frac{R}{2\sqrt{2}},\frac{5\pi}{4}$), ($\frac{R}{2},\pi$), ($\frac{R}{2},\frac{\pi}{2}$).

\subsection{Baseline methods}\label{Baseline methods}
In this study, we selected five representative state-of-the-art methods:
\begin{itemize}
    \item \textbf{STnet \cite{STnet}} utilizes DenseNet-121 as the image encoder to extract H\&E image features, which were then embedded into the feature space and projected onto the dimension of gene expression through fully connected layers.
    \item \textbf{DeepSpaCE \cite{DeepSpaCE}} operates similarly to STNet, but it employs the VGG16 network for gene expression prediction.
    \item \textbf{HisToGene \cite{HisToGene}} adopts a vision Transformer as the image encoder, leveraging self-attention mechanism to extract global features, which were subsequently projected onto the dimension of gene expression through fully connected layers.
    \item \textbf{THItoGene \cite{ThItogene}} uses H\&E images as input and employed dynamic convolutional and capsule networks to capture signals of potential molecular features within histological samples.
    \item \textbf{STAGE \cite{STAGE}} utilizes a spatially supervised autoencoder generator to produce the two-dimensional or three-dimensional coordinates of unmeasured points, which are then used to generate the corresponding high resolution gene expression data.
\end{itemize}

\subsection{Implementation Details}
For all baselines, we used the default parameters specified in the original papers.
Our experiments were executed on a single NVIDIA RTX 4090 GPU using PyTorch (version 2.2.0) and Python 3.10. The training protocol was established for 1000 epochs, entailinga batch size of 512 and a learning rate set at 0.001.

\subsection{HisToSGE enables more accurate generation of high-resolution gene expression.}
To quantitatively evaluate the performance of HisToSGE and other methods in generating high-density gene expression profiles across all datasets, we randomly removed 50\% of the spots, using the remaining spots as the training set and the removed spots as the test set. The generation performance was assessed by comparing the PCC, MSE, and MAE between the original data and the recovered data from different methods. As shown in Table \ref{tab2}, HisToSGE achieved the highest PCC and the lowest MSE and MAE across all datasets. In evaluating tissue slices from 151676 of DLPFC, Mouse Brain, BC1, and BC2, the HisToSGE method demonstrates superior performance compared to the second-best method, STAGE. HisToSGE achieves PCC that are 27\%, 9\%, 32\%, and 15\% higher than those of STAGE for the respective tissue slices. In terms of error reduction, HisToSGE demonstrates MSE that are 14\%, 11\%, 11\%, and 4\% lower. Similarly, the method also achieves MAE that are 28\%, 14\%, 13\%, and 9\% lower compared to STAGE.\par
To visually analyze the performance of different methods in generating high-density gene expression profiles, we conducted spatial visualization of the partially generated data and compared it with the downsampled data and the true expression. We presented the gene expression generation results across four datasets using various methods (\autoref{fig2} and \autoref{fig3}). Compared to other methods, HisToSGE's generation results are closer to the true gene expression patterns, demonstrating superior performance. As shown in \autoref{fig2}A, for the marker genes \textit{PCP4}, \textit{FABP4}, and \textit{MBP} on the 151676 slice from the DLPFC dataset, the PCC values between HisToSGE's generated gene expression and the true gene expression are 0.71, 0.72, and 0.85, respectively, which are 21\%, 17\%, and 7\% higher than those achieved by the second-best method, STAGE. Additionally, as shown in \autoref{fig2}B and \autoref{fig3}, HisToSGE also demonstrates excellent performance in gene expression generation on the Mouse Brain, BC1, and BC2 datasets. In contrast, DeepSpaCE and STNet showed relatively poor performance, failing to effectively retain the original gene expression patterns.

\subsection{HisToSGE enhances gene expression patterns}
Here, we downsampled the low-resolution histological images by a factor of N and used these images as the training set to generate high-resolution gene expression data. We presented the generated high-resolution gene expression data with downsampling factors ranging from 2x to 8x. We compared the raw and generated gene expression profiles of several marker genes, including \textit{MOBP}, \textit{SCGB2A2}, \textit{HBB}, and \textit{PCP4} (\autoref{fig4}), which are highly expressed in myelin. Compared to the raw data, the generated gene expression profiles displayed clearer expression patterns, demonstrating the effective enhancement capability of HisToSGE. 

\subsection{HisToSGE can better preserve spatial structures}
To further explore the effectiveness of HisToSGE in spatial domain recognition tasks, we conducted clustering analysis on both the original and generated data using K-means, STAGATE \cite{STGATE}, and STMask \cite{STMask} clustering methods. Manual annotations served as the ground truth, and we evaluated the results using the ARI as the evaluation metric.\par
In spatial domain recognition, HisToSGE showed improvements over STAGE and THItoGene in K-means (ARI scores for the recovered data of HisToSGE, STAGE, and THItoGene were 0.37, 0.31, and 0.28, respectively), STAGATE (ARI scores were 0.52, 0.50, and 0.44, respectively), and STMask (ARI scores were 0.54, 0.51, and 0.42, respectively) (\autoref{fig5} A, B and C). \par
Additionally, we visualized the clustering results of slices 151673 and 151674 from the DLPFC datasets using K-means, STAGATE, and STMask methods. The clustering results in \autoref{fig5}D demonstrate that HisToSGE consistently outperforms other methods across different slices and clustering techniques. In slice 151673, HisToSGE achieves higher ARI scores than STAGE and THItoGene across K-means, STAGATE, and STMask clustering methods, with particularly notable performance in the STMask clustering. This trend is also observed in slice 151674, where HisToSGE consistently achieves higher ARI scores than the other methods in all clustering approaches. These results indicate that HisToSGE is more accurate and robust in spatial domain recognition, effectively capturing the spatial structure of the data. 

\subsection{Ablation studies}
To evaluate the learning capabilities of various feature extraction backbone networks for image features, we conducted ablation experiments (\autoref{tab3}). We tested three different backbone networks, and the results indicated that the Transformer network achieved the highest performance. Furthermore, the FeedForward backbone network surpassed the performance of the second-ranked method, STAGE, suggesting that rich histological image features can effectively represent gene expression profiles.\par
We also developed the STAGE\_Plus model, which utilizes rich image features extracted by the UNI module to replace gene expression features as the input for STAGE. In this model, STAGE uses these image features to generate two-dimensional coordinates of spots through an encoder, which are subsequently employed to generate gene expression profiles. The findings suggest that two-dimensional coordinates alone are insufficient to represent the rich image and gene expression features. In contrast, the feature learning module of HisToSGE can effectively learn these rich image features, enabling high-resolution gene expression mapping. 

\section{CONCLUSIONS}
The generation of high-resolution ST is a pivotal challenge in both experimental and computational biology. In this study, we developed a robust method, HisToSGE, to integrate histological images, gene expression data, and spatial location information. 
This method aims to learn rich image features and continuous spatial expression patterns, and predict spatial gene expression profiles for unmeasured spots at any resolution. HisToSGE consists of two main modules: the feature extraction module and the feature learning module. The feature extraction module generates rich image features and combines multimodal image features. The feature learning module employs a multi-head attention mechanism to integrate spot coordinates and learn features from multimodal feature maps, thereby enhancing feature representation.
We compared our method with other approaches on multiple real ST datasets. The results demonstrate that, in generating high-density gene expression profiles, our method improves the average PCC by 9\% to 32\% compared to state-of-the-art methods. Additionally, our method not only enhances gene expression patterns but also effectively preserves the original spatial structure.

\section*{Acknowledgment}
The work was supported in part by the National Natural Science Foundation of China (62262069), in part by the Program of Yunnan Key Laboratory of Intelligent Systems and Computing (202205AG070003), and the Yunnan Talent Development Program - Youth Talent Project. 

\small
\bibliography{Ref.bib} 

\begin{thebibliography}{10}
\providecommand{\url}[1]{#1}
\csname url@samestyle\endcsname
\providecommand{\newblock}{\relax}
\providecommand{\bibinfo}[2]{#2}
\providecommand{\BIBentrySTDinterwordspacing}{\spaceskip=0pt\relax}
\providecommand{\BIBentryALTinterwordstretchfactor}{4}
\providecommand{\BIBentryALTinterwordspacing}{\spaceskip=\fontdimen2\font plus
\BIBentryALTinterwordstretchfactor\fontdimen3\font minus \fontdimen4\font\relax}
\providecommand{\BIBforeignlanguage}[2]{{%
\expandafter\ifx\csname l@#1\endcsname\relax
\typeout{** WARNING: IEEEtran.bst: No hyphenation pattern has been}%
\typeout{** loaded for the language `#1'. Using the pattern for}%
\typeout{** the default language instead.}%
\else
\language=\csname l@#1\endcsname
\fi
#2}}
\providecommand{\BIBdecl}{\relax}
\BIBdecl

\bibitem{rao2021exploring}
A.~Rao, D.~Barkley, G.~S. Fran{\c{c}}a, and I.~Yanai, ``Exploring tissue architecture using spatial transcriptomics,'' \emph{Nature}, vol. 596, no. 7871, pp. 211--220, 2021.

\bibitem{chen2022spatiotemporal}
A.~Chen, S.~Liao, M.~Cheng, K.~Ma, L.~Wu, Y.~Lai, X.~Qiu, J.~Yang, J.~Xu, S.~Hao \emph{et~al.}, ``Spatiotemporal transcriptomic atlas of mouse organogenesis using dna nanoball-patterned arrays,'' \emph{Cell}, vol. 185, no.~10, pp. 1777--1792, 2022.

\bibitem{spacel}
H.~Xu, S.~Wang, M.~Fang, S.~Luo, C.~Chen, S.~Wan, R.~Wang, M.~Tang, T.~Xue, B.~Li \emph{et~al.}, ``Spacel: deep learning-based characterization of spatial transcriptome architectures,'' \emph{Nature Communications}, vol.~14, no.~1, pp. 7603--7621, 2023.

\bibitem{chen2020spatial}
W.-T. Chen, A.~Lu, K.~Craessaerts, B.~Pavie, C.~S. Frigerio, N.~Corthout, X.~Qian, J.~Lal{\'a}kov{\'a}, M.~K{\"u}hnemund, I.~Voytyuk \emph{et~al.}, ``Spatial transcriptomics and in situ sequencing to study alzheimer’s disease,'' \emph{Cell}, vol. 182, no.~4, pp. 976--991, 2020.

\bibitem{cui2023spatiotemporal}
G.~Cui, K.~Dong, J.-Y. Zhou, S.~Li, Y.~Wu, Q.~Han, B.~Yao, Q.~Shen, Y.-L. Zhao, Y.~Yang \emph{et~al.}, ``Spatiotemporal transcriptomic atlas reveals the dynamic characteristics and key regulators of planarian regeneration,'' \emph{Nature Communications}, vol.~14, no.~1, pp. 3205--3221, 2023.

\bibitem{HisToGene}
M.~Pang, K.~Su, and M.~Li, ``Leveraging information in spatial transcriptomics to predict super-resolution gene expression from histology images in tumors,'' \emph{BioRxiv}, pp. 1--31, 2021.

\bibitem{iStar}
D.~Zhang, A.~Schroeder \emph{et~al.}, ``Inferring super-resolution tissue architecture by integrating spatial transcriptomics with histology,'' \emph{Nature Biotechnology}, pp. 1--9, 2024.

\bibitem{bayes}
E.~Zhao, M.~R. Stone \emph{et~al.}, ``Spatial transcriptomics at subspot resolution with bayesspace,'' \emph{Nature Biotechnology}, vol.~39, no.~11, pp. 1375--1384, 2021.

\bibitem{xue2024stentrans}
S.~Xue, F.~Zhu, C.~Wang, and W.~Min, ``stentrans: Transformer-based deep learning for spatial transcriptomics enhancement,'' in \emph{International Symposium on Bioinformatics Research and Applications}.\hskip 1em plus 0.5em minus 0.4em\relax Springer, 2024, pp. 63--75.

\bibitem{wsi1}
L.~N. Waylen, H.~T. Nim, L.~G. Martelotto, and M.~Ramialison, ``From whole-mount to single-cell spatial assessment of gene expression in 3d,'' \emph{Communications Biology}, vol.~3, no.~1, pp. 602--613, 2020.

\bibitem{mclSTExp}
W.~Min, Z.~Shi, J.~Zhang, J.~Wan, and C.~Wang, ``Multimodal contrastive learning for spatial gene expression prediction using histology images,'' \emph{arXiv preprint arXiv:2407.08216}, pp. 1--9, 2024.

\bibitem{STnet}
B.~He, L.~Bergenstr{\aa}hle, L.~Stenbeck, A.~Abid, A.~Andersson, {\AA}.~Borg, J.~Maaskola, J.~Lundeberg, and J.~Zou, ``Integrating spatial gene expression and breast tumour morphology via deep learning,'' \emph{Nature Biomedical Engineering}, vol.~4, no.~8, pp. 827--834, 2020.

\bibitem{DeepSpaCE}
T.~Monjo, M.~Koido, S.~Nagasawa, Y.~Suzuki, and Y.~Kamatani, ``Efficient prediction of a spatial transcriptomics profile better characterizes breast cancer tissue sections without costly experimentation,'' \emph{Scientific Reports}, vol.~12, no.~1, pp. 4133--4145, 2022.

\bibitem{ViT}
A.~Dosovitskiy, L.~Beyer \emph{et~al.}, ``An image is worth 16x16 words: Transformers for image recognition at scale,'' in \emph{International Conference on Learning Representations}, 2021, pp. 1--22.

\bibitem{ThItogene}
Y.~Jia, J.~Liu, L.~Chen, T.~Zhao, and Y.~Wang, ``{THItoGene}: a deep learning method for predicting spatial transcriptomics from histological images,'' \emph{Briefings in Bioinformatics}, vol.~25, no.~1, pp. 464--474, 2024.

\bibitem{STAGE}
S.~Li, K.~Gai, K.~Dong, Y.~Zhang, and S.~Zhang, ``High-density generation of spatial transcriptomics with stage,'' \emph{Nucleic Acids Research}, vol.~52, no.~9, pp. 4843--4856, 2024.

\bibitem{UNI}
R.~J. Chen, T.~Ding, M.~Y. Lu, D.~F. Williamson, G.~Jaume, A.~H. Song, B.~Chen, A.~Zhang, D.~Shao, M.~Shaban \emph{et~al.}, ``Towards a general-purpose foundation model for computational pathology,'' \emph{Nature Medicine}, vol.~30, no.~3, pp. 850--862, 2024.

\bibitem{DLPFC}
K.~R. Maynard, L.~Collado-Torres \emph{et~al.}, ``Transcriptome-scale spatial gene expression in the human dorsolateral prefrontal cortex,'' \emph{Nature Neuroscience}, vol.~24, no.~3, pp. 425--436, 2021.

\bibitem{MouseBrain}
S.~M. Sunkin, L.~Ng, C.~Lau, T.~Dolbeare, T.~L. Gilbert, C.~L. Thompson, M.~Hawrylycz, and C.~Dang, ``Allen brain atlas: an integrated spatio-temporal portal for exploring the central nervous system,'' \emph{Nucleic Acids Research}, vol.~41, no.~D1, pp. D996--D1008, 2012.

\bibitem{10x}
A.~Janesick, R.~Shelansky \emph{et~al.}, ``High resolution mapping of the tumor microenvironment using integrated single-cell, spatial and in situ analysis,'' \emph{Nature Communications}, vol.~14, no.~1, pp. 8353--8368, 2023.

\bibitem{STGATE}
K.~Dong and S.~Zhang, ``Deciphering spatial domains from spatially resolved transcriptomics with an adaptive graph attention auto-encoder,'' \emph{Nature Communications}, vol.~13, no.~1, pp. 1739--1751, 2022.

\bibitem{STMask}
W.~Min, D.~Fang, J.~Chen, and S.~Zhang, ``Dimensionality reduction and denoising of spatial transcriptomics data using dual-channel masked graph autoencoder,'' \emph{bioRxiv}, pp. 01--20, 2024.

\end{thebibliography}
\bibliographystyle{IEEEtran}  
\end{document}